# SPECTRAL DENSITY OF PHASE NOISE INTER-LABORATORY COMPARISON FINAL RESULTS

P. Salzenstein[1], J. Čermak[2], R. Barillet[3], F. Lefebvre[4], W. Schaefer[5], G. Cibiel[6], G. Sauvage[7], O. Franquet[8], O. Llopis[9], F. Meyer[10], N. Franquet[1], A. Kuna[2], L. Šojdr[2], and G. Hejc[5]

[1] FEMTO-ST, CNRS, Associated to LNE, 32 av. Obs., F25044 Besançon Cedex, France
Phone: +33(0)381853999 Fax: +33(0)381853998
[2] Czech Academy of Science, Dpt of Standard Time & Frequency, Chaberska 57, 18251 Praha 8, Czech Republic
[3] Observatoire de Paris, CNRS, LNE-SYRTE, Bât B, 61 av. Obs., F75014 Paris, France
[4] Oscilloquartz, rue des Brévards 16, CH-2002 Neuchâtel 2, Switzerland
[5] TimeTech GmbH., Curiestrasse 2, D70563 Stuttgart, Germany
[6] CNES (Centre national d'Etudes Spatiales), 18 av. E. Belin, F31404 Toulouse Cedex France
[7] AEROFLEX Test Solutions, 5 Place du Général De Gaulle, F78990 Elancourt, France
[8] A.R. Electronique, rue du Bois de la Courbe, F25870 Châtillon-le-Duc, France
[9] LAAS CNRS, 7 av. du Colonel Roche, F31077 Toulouse France
[10] Observatoire de Besançon, CNRS, Associated to LNE, 41bis av. de l'Obs., BP1615, F25010 Besançon, France

## Résumé

Ce papier rapporte les principaux résultats de la comparaison en mesures de densité spectrales de bruit de phase réalisée d'octobre 2005 à décembre 2006 sur deux paires d'oscillateurs à 5 et 100 MHz ainsi qu'un ORD à 3.5 GHz. Il ne s'agit pas tant de comparer les performances des oscillateurs, mais bien de comparer et évaluer les incertitudes et la résolution des bancs ainsi que la reproductibilité des mesures. Le but n'est pas de faire une compétition entre les différents moyens de mesures existants mais plutôt de raccorder plusieurs systèmes afin d'améliorer la confiance que l'on peut avoir dans les mesures de bruit de phase.

## Abstract

This paper reports main results of the phase noise comparison that has been performed between october 2005 and december 2006, using two oscillators at 5 and 100 MHz and un DRO at 3.5 GHz. The problem is not to compare the performances of several oscillators, but to compare and to make an evaluation of the uncertainties, and of course the resolution and the reproducibility of the measurements. This comparison allow us to determine the ability to get various systems traceable together in order to increase the trust that one can have in phase noise measurements.

## Introduction

Une comparaison internationale en bruit de phase a été organisée en 1993 et publiée lors de l'EFTF de 1994 [1]. Plus d'une dizaine d'année après, il convient de tester les nouveaux bancs de laboratoire réalisés en interne ou commerciaux. A la demande du LNE (le Bureau National de Métrologie - BNM - fait partie intégrante du LNE depuis le début de l'année 2005), l'institut FEMTO-ST, en tant que Laboratoire Associé au LNE sous le numéro d'accréditation COFRAC numéro 2.13, est chargé d'organiser une comparaison en bruit de phase portant sur des mesures à 5 MHz, à 100 MHz et en hyperfréquences. Il s'agit de comparer et évaluer les incertitudes d'une part, et la résolution des bancs et la reproductibilité des mesures, ce qui intéresse notamment les fabricants de bancs de mesures de bruit de phase.

Les oscillateurs à caractériser en bruit de phase au cours de cette comparaison sont des oscillateurs commerciaux. Ils seront prêtés par la société OSCILLOQUARTZ et par FEMTO-ST en ce qui concerne les oscillateurs à 5 MHz, et FEMTO-ST fournit des oscillateurs AR ELECTRONIQUE à 100 MHz. Un Oscillateur à Résonateur Diélectrique (ORD) commercial prêté par le LAAS-CNRS est utilisé pour la caractérisation hyperfréquence à 3,5 GHz.

## Généralités sur les mesures spectrales

Il est possible de caractériser les instabilités de fréquence dans le domaine fréquentiel par l'étude du spectre ou dans le domaine temporel par l'étude statistique des différents résultats de comptage de la fréquence de ce signal. La densité spectrale de bruit de phase est définie en intégrant le rapport bandes latérales sur porteuse donné en fonction de l'écart à la porteuse, c'est-à-dire en fonction des fréquences de Fourier. Dans le domaine temporel, la variance dite d'Allan résulte de l'étude statistique des résultats de comptages de la fréquence, et permet de caractériser l'instabilité de l'oscillateur en fonction du temps de comptage. Un oscillateur présente généralement près de la porteuse un bruit Flicker de fréquence, en 1/f dans le domaine des fluctuations de fréquence alias en $1/f^3$ pour la densité spectrale de bruit de phase, qui procure un palier dit Flicker pour la variance d'Allan dans le domaine temporel. Attention, la correspondance n'est pas bijective: un bruit de phase en $1/f^3$ donne un palier en variance d'Allan, l'inverse n'est pas forcément vrai.

Le principe des mesures spectrales consiste à démoduler en phase le signal à étudier en asservissant l'oscillateur à étudier sur un signal de référence, au moyen d'une boucle de phase. En particulier, en hyperfréquence, on asservit le DUT et non la référence, qui est déjà asservie sur un OUS basse fréquence.

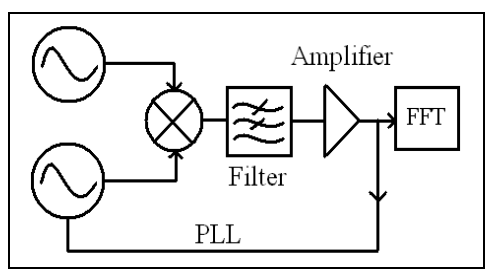

La tension d'erreur de cet asservissement est alors proportionnelle à l'écart de phase entre l'oscillateur sous test et la référence. En dehors de la bande d'asservissement, elle est donc proportionnelle à l'écart de phase entre l'oscillateur sous test (free-running) et la référence. Elle est proportionnelle aux fluctuations de fréquence dans la bande passante de l'asservissement.

Toutefois, ce n'est pas toujours vrai semble-t-il, car cela dépend des caractéristiques détaillées de la PLL, en particulier du nombre d'intégrations dans la FT de l'asservissement : c'est vrai s'il y a une seule intégration, précisément celle apportée par la comparaison de phase, complétée par un ampli simple; mais pour des oscillateurs bruyants, ou en hyperfréquence, on est souvent obligé de mettre deux intégrations avec une avance de phase, et là ce n'est plus vrai sauf près de la fréquence de coupure de la PLL.

Les résultats dans la bande passante d'asservissement ne sont pas à utiliser systématiquement, car ils dépendent beaucoup trop des caractéristiques de la PLL. Si l'on connaît bien les caractéristiques de la PLL, on peut estimer les changements de pente pour le bruit du DUT, bien que cela semble audacieux, en tout cas la qualité métrologique devient discutable, sauf à avoir très bien caractérisé la PLL. Cette tension est amplifiée et envoyée sur un analyseur de spectre dynamique qui calcule et affiche sa densité spectrale des fluctuations.

## Reproductibilité des mesures

Les spectres obtenus seront susceptibles d'être affectées par l'environnement des mesures, à savoir les conditions de température, d'hygrométrie, d'exposition à des rayonnements électromagnétiques, d'alimentation des oscillateurs à tester et des instruments de mesures, mais aussi liées aux conditions de démarrage des oscillateurs et à leur nature, au temps de stockage après allumage, et aux bancs eux-mêmes.

Il s'agit donc d'être capable d'évaluer les différentes contributions possibles. Ces objectifs sont ambitieux aussi est-il envisagé de suivre préférentiellement la reproductibilité inter-laboratoire sur des bancs de même nature ou utilisant des méthodes équivalentes, mais aussi des mesures effectuées avec des méthodes variées, tout en restant dans le contexte de mesures effectuées en laboratoire. Comme évoqué dans l'introduction, le but n'est pas tant de comparer les performances des oscillateurs, mais de comparer et d'évaluer d'une part les incertitudes, et d'autre part la résolution des bancs et la reproductibilité des mesures.

## Objectif général de l'étude

Cette comparaison inter-laboratoire en bruit de phase a pour objectif d'évaluer la reproductibilité des mesures effectuées par les différents bancs utilisés au sein de laboratoires de métrologie.

Etalons dits "voyageurs":

Paire d'oscillateurs à 5 MHz : oscillateurs BVA de fabrication Oscilloquartz, de type 8607 fournis par Oscilloquartz (ref. 102) et FEMTO-ST (ref. 172)

Paire d'oscillateurs à 100 MHz : oscillateurs Jumbostar de fabrication AR Electronique, de type AR1492, numéros 1 et 2, fournis par FEMTO-ST

ORD hyperfréquence à 3,5 GHz, de fabrication MITEQ, de type DRO D03500, numéro de série 613318 fourni par le LAAS-CNRS

## Protocole de mesure

Mesure préliminaires

Les oscillateurs ou paires d'oscillateurs seront mesurés par le laboratoire organisant la comparaison en début et en fin de comparaison.

Réception des étalons voyageurs

Au moment de la réception des colis, il est précisé l'état de l'emballage et des oscillateurs, ainsi que toute information utile. Ces derniers seront immédiatement mis en chauffe 48 heures dans le local approprié avant d'effectuer les mesures. La fréquence et la puissance délivrée sont vérifiés préalablement. Il est à noter que la stabilité est affectée par la durée du temps de coupure: de manière empirique, prévoir un temps d'attente double du temps de coupure. Pour mémoire, les performances spécifiées par les fabricants sont garanties à l'issue d'une période de 90 jours sans interruption. Toutefois, il n'est pas pensable dans le cadre de la circulation des étalons, de respecter ce délai. De plus les oscillateurs sont sensibles aux chocs.

Mesures

Les oscillateurs ou paires d'oscillateurs seront mesurés en terme de spectre de densité spectrale de puissance de bruit de phase en fonction des fréquences de Fourier comprises entre 1 Hz et 100 kHz.

Attention toutefois : en hyperfréquences, on ne pourra pas mesurer à 1Hz pour un ORD car les fluctuations en free-running y sont de plusieurs $rad^2/Hz$. Aussi est-il préférable de n'indiquer cela que pour les mesures à 5MHz et 100MHz. Pour l'ORD, il vaut mieux commencer à 1kHz environ. Sont notamment données les valeurs aux fréquences de Fourier suivantes 1 Hz, 10 Hz, 100 Hz, 1 kHz, 10 kHz, 100 kHz auxquelles des incertitudes à 2 $\sigma$ seront associées. Des résultats de mesures donnés sous la forme d'un certificat d'étalonnage sont appréciées dans la mesure du possible, certains laboratoires participant à cette comparaison ne réalisant pas normalement d'étalonnage, accrédités ou non.

Il est intéressant que les laboratoires disposant de plusieurs bancs de mesures puissent les utiliser pour comparer divers types de bancs de mesures de bruit de phase.

Les bancs de mesures de bruit de phase potentiellement identifiés sont les suivant :

- bancs Hewlett Packard
- bancs Europtest
- bancs Femtosecond
- bancs Timing Solutions
- bancs AR Electronique

D'autre part, certains laboratoires disposent de moyens de mesures de stabilité court terme de fréquence permettant de déduire l'allure du spectre de phase, et qui peuvent donner des informations intéressantes.

Afin que les mesures effectuées au sein des différents laboratoires participants ne soient pas trop affectées par l'environnement, il est intéressant de préciser les conditions des mesures ambiantes. Il s'agit notamment des paramètres suivants : la température du local, l'hygrométrie, mais aussi de préciser éventuellement si le banc de mesures et les oscillateurs sont placés dans une enceinte type cage de Faraday, et la nature des alimentations utilisées, par exemple, batteries pour les oscillateurs ou secteur pour les instruments composant le banc. Il sera précisé si les incertitudes associées aux mesures sont calculées en fonction de ces paramètres ou données dans le cadre d'accréditations.

Réexpédition des étalons voyageurs

Les oscillateurs sont ensuite emballés et réexpédiés au laboratoire devant effectuer les mesures suivantes en respectant l'ordre prévu, sauf s'il est précisé qu'il doit y avoir une inversion par le responsable de la comparaison. Toute information jugée utile est transmise au responsable de la comparaison.

Transmission des résultats

Les résultats complets seront transmis au responsable de la comparaison avant que les étalons voyageurs aient fini de circuler entre les laboratoires participants à la comparaison.

Il sera précisé les conditions ambiantes des mesures effectuées comme listées plus haut, le type de banc et la méthode utilisée ainsi que les incertitudes associées.

Chaque laboratoire participant est codifié par une lettre afin de garantir l'anonymat de chaque participant.

# Résultats

Les résultats obtenus concernent tout d'abord les niveau de densité spectrale de bruit de phase mesurés à 5 MHz, 100 MHz et 3,5 GHz [2]. La comparaison donne également des résultats interessants concernant les bancs eux-mêmes.

| dBc/Hz | $10^0$ Hz | $10^1$ Hz | $10^2$ Hz | $10^3$ Hz | $10^4$ Hz | $10^5$ Hz |
|---|---|---|---|---|---|---|
| LR 1 | -125.5 ±2 | -145 ±2 | -151.5 ±2 | -156 ±2 | -154 ±2 | -156 ±2 |
| LR 2 | -125±2 | -136 ±2 | -140 ±2 | -154 ±2 | -154 ±2 | -155 ±2 |
| A | -126±2 | -145 ±2 | -151.5 ±2 | -155 ±2 | -155 ±2 | -155.5 ±2 |
| B | -113 ±5 | -135 ±5 | -143 ±5 | -149 ±5 | -155 ±5 | -157.5 ±5 |
| C* | -126 | -145.5 | -151.5 | -155 | -155.5 | -156.5 |
| D* | -125.5 | -145.5 | -152 | -156 | -155.5 | -156.5 |
| E |  | -144 ±2 | -154 ±2 | -158 ±2 | -158 ±2 | -159 ±2 |
| F* | -126 |  |  | -155 | -155 | -155 |
| G | -126 ±2 | -144.5 ±2 | -151.5 ±2 | -155.5 ±2 | -155.5 ±2 | -156 ±2 |
| H | -126.08 ±3 | -145.30 ±3 | -152.08 ±3 | -155.57 ±3 | -155.50 ±3 | -157.59 ±3 |
| I | -122.5 ±3 | -142 ±3 | -149 ±3 | -154±3 |  |  |
| LR 2 | -124.5 ±2 | -142 ±2 | -148 ±2 | -154±2 | -154 ±2 | -155.8 ±2 |
| LR 1 | -125.4 ±2 | -144 ±2 | -151 ±2 | -156 ±2 | -154 ±2 | -156 .8 ±2 |

Table 1: Niveau de bruit (en dBc/Hz) en fonction de la fréquence de Fourrier pour une porteuse de 5 MHz pour chaque laboratoire codifié par une lettre. Les incertitudes sont données à 2σ

La table 1 représente le résultats des mesures sans aucune correction pour la paire d'oscilateurs. Il n'a pas été considéré que le poids des deux oscilateurs était le même.

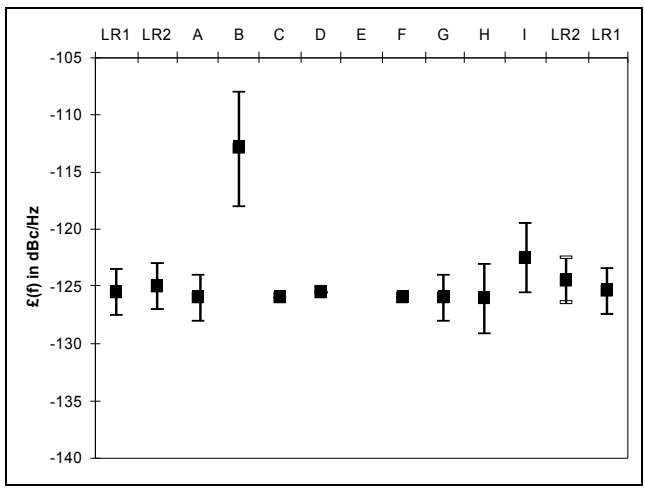

Mesures à 1 Hz pour une porteuse à 5 MHz. Incertitudes sont données à 2σ

Les résultats sont présentés sur la figure ci-dessus. A 5 MHz, un des laboratoires participants a eu un problème qui reste encore à préciser. Trois autres laboratoires n'ont pas fait parvenir leur barres d'incertitudes, ce qui explique que l'on n'ait pas indiqué les barres correspondantes. Le niveau de bruit semble être assez semblable pour la plupart des laboratoires participants. Bien qu'il reste encore un offset pas encore expliqué pour un des participans. Sur les figures également les laboratoires sont codifiés par des lettres.

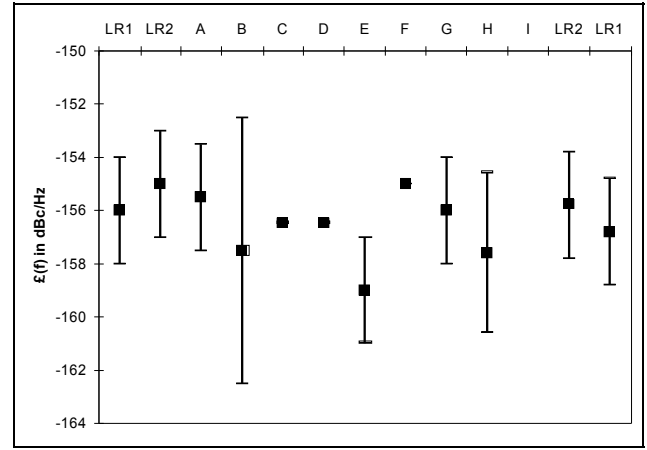

Mesures à 100 kHz pour une porteuse à 5 MHz. Incertitudes sont données à 2σ

| dBc/Hz | $10^0$ Hz | $10^1$ Hz | $10^2$ Hz | $10^3$ Hz | $10^4$ Hz | $10^5$ Hz |
|---|---|---|---|---|---|---|
| LR 1 | -64 ±2 | -96 ±2 | -131 ±2 | -153 ±2 | -161 ±2 | -162 ±2 |
| LR 2 | -58 ±2 | -98 ±2 | -129 ±2 | -155 ±2 | -162 ±2 | -162 ±2 |
| A | -64 ±2 | -95.5 ±2 | -129.5 ±2 | -153.5 ±2 | -160 ±2 | -160 ±2 |
| B | -65 ±3 | -100 ±3 | -133.5 ±3 | -152 ±3 | -160 ±3 | -161 ±3 |
| C* | -70 | -102 | -134 | -155 | -163 | -163 |
| D* | -73 | -100 | -131 | -156.5 | -162 | -162.5 |
| E |  |  | -135 ±2 | -151 ±2 | -158 ±2 | -159 ±2 |
| F* | -68 |  | -128 |  | -163 | -163 |
| G | -67 ±2 | -97 ±2 | -129.5 ±2 | -153.5 ±2 | -160.5 ±2 | -160.5 ±2 |
| H | -76.3 ±3 | -96 ±3 | -130.5 ±3 | -154.8 ±3 | -161.6 ±3 | -161.7 ±3 |
| I | -86 | -100 | -129 | -150 | -158 |  |
| LR 2 | -60±2 | -95 ±2 | -130 ±2 | -153 ±2 | -161 ±2 | -161 ±2 |

Table 2: Niveau de bruit (en dBc/Hz) en fonction de la fréquence de Fourrier pour une porteuse de 100 MHz pour chaque laboratoire codifié par une lettre. Les incertitudes sont données à 2σ

Les résultats des mesures à 100 MHz sont présentés dans ce qui suit. Les résultats sont exprimés en dBc/Hz sans aucune correction comme dans le cas précédent.

Les mesures semblent cohérentes. Cependant, près de la porteuse à 100 MHz, des problèmes liés à la stabilité ont rendues les mesures plus difficiles surtout à 1 Hz. Des différences notables existent pour certains des participants et une investigation plus approfondie doit être menée. En ce qui concerne les mesures de bruit de phase pour des quartz à 100 MHz, il est classiquement demandé les valeurs à 100 Hz de la porteuse, que nous avons également choisi de représenter sur les figures. Les résultats sont présentés en dBc/Hz.

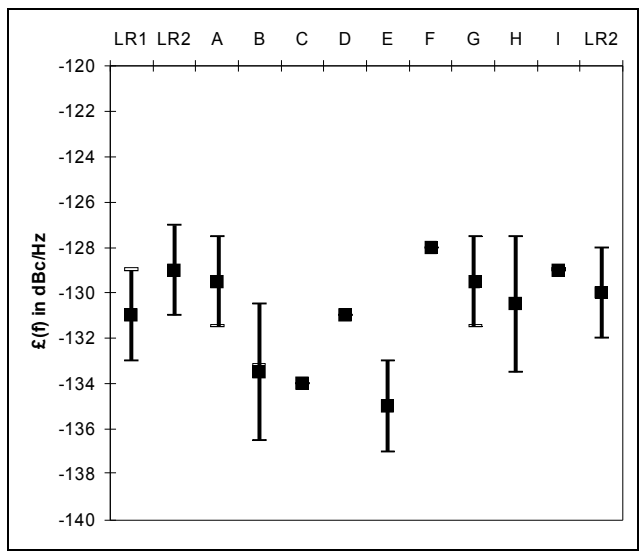

Mesures à 100 Hz pour une porteuse à 100 MHz.
Incertitudes sont données à 2σ

Pour un des participants, le plancher de bruit n'a pas été mesuré à 100 kHz. L'incertitude n'a pas été donnée pour quatre des laboratoires participants, pour lesquels nous n'avons pas indiqué de barre d'incertitude.

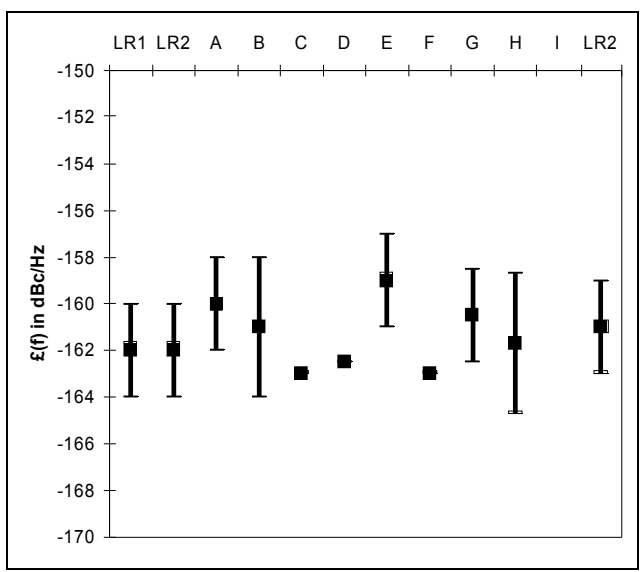

Mesures à 100 kHz pour une porteuse à 100 MHz.
Incertitudes sont données à 2σ

Le DRO à 3.5 GHz a été mesuré seulement dans deux laboratoires. Les résultats ne sont donc pas suffisamment représentatifs pour constituer une comparaison proprement dite. Toutefois, des informations intéressantes ont pu être dégagées. En effet, les mesures effectuées à 1 kHz differaient. Les premières série de mesures effectuées dans un des laboratoires étaient trop optimistes. Ces différences ont pu être levées lorsque des isolateurs ont été intercalés. A 100 Hz de la porteuse, les mesures sur l'ORD dans les deux laboratoires sont de -57,8±2 et de -60,3±3 dB.rad²/Hz. A 10 kHz, les résultats de mesures sont respectivement de -112,4±2 et de -116,82±3 dB.rad²/Hz. Ce résultat cohérent reste cependant non significatif en raison du faible nombre de participant à la mesure de cet oscillateur.

Le résultats des mesures à 5 et 100 MHz reste comparable et relativement proches; Les différences ne peuvent s'expliquer qu'à l'aide d'une analyse fine des différents paramètres et de la manière dont chaque laboratoire a mené sa campagne de mesures. Lorsque l'écart excède 2 dB, il est toutefois nécessaire de rechercher plus avant les causes possibles des écarts mesurés.

Discussion:

Les conditions environementales sont précisées dans la table suivante. Dans ce tableau ont été rapportés diffrents paramètres comme la température et l'hygrométrie. Il est également précisé si les oscillateurs étaient ou non branchés sur batterie ou sur secteur. Précisons qu'il n'a pas été possible de les laisser toujours sur batterie lors des transports, notamment lors des déplacement en avion, c'est pourquoi la décision d'utiliser des batteries a été laissée aux soins des laboratoires. Certaines des mesures ont été réalisées en cage de Faraday pour protéger les oscillateurs des rayonnements électromagnétiques.

|  | Temperature (°C) | hygrometrie (%) | Batteries | Cage de Faraday |
|---|---|---|---|---|
| LR 1 | 21.5 ±2.5 | ? | NON | NON |
| LR 2 | 25 ±2 | 36 ±5 | NON | NON |
| A | 23 | 44 | OUI | OUI |
| B | 22.5 ±1.5 | ? | NON | NON |
| C | 23.5 ±2.5 | 42.5 ±12.5 | OUI | NON |
| D | 20.5 ±2 | ? | OUI | NON |
| E | 23 ±3 | ? | OUI | OUI |
| F | temperature de la pièce | ? | NON | NON |
| G | 21 ±1 | ? |  |  |
| H | 23 ±1 | 16 ±5 | OUI | OUI |
| I | 22 ±1 | ? | NON | NON |
| LR 2 | 26±2 | 42 ±5 | NON | NON |

Table 3: conditions environementales de smesures

Les conditions dans lesquelles les oscillateurs ont été mis le cas échéant sous tension et démarrés, ne sont pas encore précisées. Comme les oscillateurs sont éteints durant le transport, cela peut avoir un impact sur la stabilité et sur le niveau de bruit de phase à 1 Hz de la porteuse. Toutes les données n'ont pas pu être collectées.

La reproductibilité des mesures semble ne pas être affectée par les conditions de températures, relativement similaires au sein des laboratoires.

Le taux d'humidité, tant qu'il est maitrisé, ne semble pas avoir d'impact dans le cadre de mesures en laboratoires.

La contribution des batteries n'apparait pas clairement lorsque l'on examine le résultat des mesures. Toutefois, nous observons sur les courbes fournies par chaque laboratoire que le niveau des raies parasites ets notablement réduit lorsque les oscillateurs sont branchés sur batterie, voire lorsque l'ensemble du banc de mesure est sur batterie. Les raies parasites du 50 Hz proviennent généralement de l'alimentation secteur ou de cables inappropriés dans le banc de mesures.

Les cages de Faraday permettent de définir une meilleure résolution pour les mesures de bruit de phase en réduisant de manière significative les raies parasites.

La combinaison de conditions environementales optimales est d'une grande aide pour de telles mesures de bruit de phase, et également pour la reproductibilité des mesures.

## **Conclusion**

Le but de la comparaison était d'évaluer la reproductibilité des résultats de mesures données par les laboratoires utilsant des bancs différents mais dans des conditions de mesures en laboratoire, c'est à dire dans des conditions métrologiques a priori comparables.

La plupart des résultats confirme que l'incertitude sur les mesures de densité spectrales de bruit de phase reste de l'ordre de ±2 dB. Bien qu'il reste des différences à expliquer pour éviter ce type de souci lors de mesures ultérieures, les résultats montrent une grande cohérence.

## **Références**